\documentclass[pra,aps,amsmath,amssymb,amsfonts,twocolumn,nofootinbib,showpacs]{revtex4}
\usepackage{bm,mathrsfs}
\usepackage{graphicx}
\usepackage{epsfig}
\usepackage{amsmath,bbm}
\usepackage{amsfonts,amssymb}
\usepackage{times}
\usepackage{verbatim}
\usepackage[sort&compress]{natbib}
\usepackage{amsmath}
\usepackage[colorlinks,breaklinks,linkcolor=blue,anchorcolor=blue,citecolor=blue,urlcolor=blue,
dvipdfm
]{hyperref}

\usepackage{array}
\newcommand{\PreserveBackslash}[1]{\let\temp=\\#1\let\\=\temp}
\newcolumntype{C}[1]{>{\PreserveBackslash\centering}p{#1}}
\begin{document}
\newtheorem{theorem}{Theorem}
\newtheorem{lemma}[theorem]{Lemma}
\newtheorem{corollary}[theorem]{Corollary}
\newtheorem{proposition}[theorem]{Proposition}
\newtheorem{definition}[theorem]{Definition}
\newtheorem{example}[theorem]{Example}
\newtheorem{conjecture}[theorem]{Conjecture}
\title[Short Title]{Ground-state blockade of Rydberg atoms and application in entanglement generation}
\author{X. Q. Shao\footnote{Corresponding author: shaoxq644@nenu.edu.cn}}
\affiliation{Center for Quantum Sciences and School of Physics, Northeast Normal University, Changchun, 130024, People's Republic of China}
\affiliation{Center for Advanced Optoelectronic Functional Materials Research, and Key Laboratory for UV Light-Emitting Materials and Technology
of Ministry of Education, Northeast Normal University, Changchun 130024, China}
\author{D. X. Li}
\affiliation{Center for Quantum Sciences and School of Physics, Northeast Normal University, Changchun, 130024, People's Republic of China}
\affiliation{Center for Advanced Optoelectronic Functional Materials Research, and Key Laboratory for UV Light-Emitting Materials and Technology
of Ministry of Education, Northeast Normal University, Changchun 130024, China}
\author{Y. Q. Ji}
\affiliation{Center for Quantum Sciences and School of Physics, Northeast Normal University, Changchun, 130024, People's Republic of China}
\affiliation{Center for Advanced Optoelectronic Functional Materials Research, and Key Laboratory for UV Light-Emitting Materials and Technology
of Ministry of Education, Northeast Normal University, Changchun 130024, China}
\author{J. H. Wu}
\affiliation{Center for Quantum Sciences and School of Physics, Northeast Normal University, Changchun, 130024, People's Republic of China}
\affiliation{Center for Advanced Optoelectronic Functional Materials Research, and Key Laboratory for UV Light-Emitting Materials and Technology
of Ministry of Education, Northeast Normal University, Changchun 130024, China}
\author{X. X. Yi}
\affiliation{Center for Quantum Sciences and School of Physics, Northeast Normal University, Changchun, 130024, People's Republic of China}
\affiliation{Center for Advanced Optoelectronic Functional Materials Research, and Key Laboratory for UV Light-Emitting Materials and Technology
of Ministry of Education, Northeast Normal University, Changchun 130024, China}

\begin{abstract}
We propose a mechanism of ground-state blockade between two $N$-type Rydberg atoms in virtue of Rydberg-antiblockade effect and Raman transition. Inspired by the quantum Zeno effect, the strong Rydberg antiblockade interaction plays a role in frequently measuring one ground state of two, leading to a blockade effect for double occupation of the corresponding quantum state. By encoding the logic qubits into the ground states, we efficiently avoid the spontaneous emission of the excited Rydberg state, and maintain the nonlinear  Rydberg-Rydberg interaction at the same time. As applications, we discuss in detail the feasibility of preparing two-atom and three-atom entanglement with ground-state blockade in closed system and open system, respectively, which shows that a high fidelity of entangled state can be obtained with current experimental parameters.

\end{abstract}
\maketitle \maketitle
\section{Introduction}\label{one}
Neutral atoms are considered as a good candidate for quantum information processing. Their stable atomic hyperfine energy states, especially suiting for encoding logic qubits, are easily controllable and measurable by making use of resonant laser pulse. On the other hand, they possess state-dependent interaction properties. When an atom is excited to the high-lying Rydberg states, the powerful dipole-dipole interaction or van der Waals interaction will significantly shift its surrounding atomic energy
levels of Rydberg states, thereby inhibiting the double or more excitations of Rydberg states, and this is the so-called Rydberg blockade phenomenon. This effect can make the atomic ensemble effectively behave as a single
two-level system, thus the idea of Jaksch {\it et al}. \cite{PhysRevLett.85.2208} for using dipolar Rydberg interactions to implement a two-qubit universal quantum gate was quickly extended to a mesoscopic
regime of many-atom ensemble qubits by Lukin {\it et al.} \cite{PhysRevLett.87.037901}. In 2009, the mechanism of Rydberg blockade was verified in experiment and two groups independently claimed that a single Rydberg-excited rubidium atom blocks excitation of a second atom set about 4 $\mu$m and 10 $\mu$m apart \cite{urban2009,gaetan2009observation}, respectively. Recently, the Rydberg blockade has been used extensively in various subfields of quantum information processing, such as quantum entanglement \cite{PhysRevLett.100.170504,PhysRevLett.102.240502,PhysRevLett.104.010502,PhysRevA.82.030306}, quantum algorithms \cite{chen2011implementation,molmer2011efficient,PhysRevA.86.042321}, quantum simulators \cite{weimer2010rydberg,PhysRevA.86.053618}, single-photon switch \cite{PhysRevLett.112.073901}, and quantum repeaters \cite{PhysRevLett.100.110506,PhysRevA.81.052311,PhysRevA.81.052329}, etc.

In contrast to the Rydberg blockade, as the shifting energy of Rydberg states is compensated by the two-photon
detuning, the effect of Rydberg antiblockade occurs, which favors a resonant two-photon transition, but counters a single-photon transition. The antiblockade in Rydberg excitation was initially predicted by Ates {\it et al.} in the two-step excitation scheme of creating an ultracold Rydberg gas \cite{PhysRevLett.98.023002}, and then observed experimentally by Amthor {\it et al.} using a time-resolved spectroscopic measurement of the Penning ionization signal \cite{PhysRevLett.104.013001}. At the aspect of quantum information processing,
the Rydberg antiblockade provides researchers with brand new ideas. Combined with asymmetric Rydberg couplings and dissipative dynamics, the Rydberg antiblockade was exploited to generate high-fidelity two-qubit Bell states and  three-dimensional entanglement \cite{PhysRevLett.111.033607,PhysRevA.89.012319}. And it is also instrumental in fast synthesis of multi-qubit logic gate \cite{shao2014one,PhysRevA.95.022319}.

 We note that a resonant excitation of Rydberg state is necessary for realizing most of Rydberg-blockade-based schemes. This requirement may cause decoherence to the system of interest due to the spontaneous emission of the excited Rydberg state, although it is considered that the
Rydberg state with a large principle quantum number has a small decay rate \cite{saffman2016quantum}. If the excited-state blockade of Rydberg atoms is replaced with a ground-state blockade, we are able to minimize the effect of atomic decay and further improve the quality of quantum information processing with Rydberg atoms. Nevertheless, the interaction of natural ground-state neutral atom is less than 1~Hz at spacings greater than 1~$\mu$m \cite{RevModPhys.82.2313}, which is unsuitable for fulfilling the blockade condition.

In this work, we put forward an efficient scheme for blocking ground states of Rydberg atoms. Our idea comes from the quantum Zeno effect \cite{PhysRevLett.86.2699,facchi2000quantum}, i.e. one can freeze the evolution of quantum system by measuring it frequently enough in its known initial state, and the same conclusion can also be made by making use of a strong continuous coupling without resorting to von Neumann's projections \cite{facchi2008quantum}. For the current scheme, the dynamical evolution of system is governed by a weak Raman coupling with strength $\Omega_{\rm eff}$. A relatively strong Rydberg antiblockade interaction with strength $\lambda$, acting as a measuring device, is used to observe the evolution of the double occupation of certain ground state. In the limit $\lambda/\Omega_{\rm eff}\gg1$, the ground-state blockade for Rydberg atoms is achieved. As its application, we will discuss in detail the prominent advantage of ground-state blockade in terms of preparing entanglement via shortcut to adiabatic
passage and quantum-jump-based feedback control, respectively.

The remainder of the paper is organized as follows. We first establish the theoretical model of ground-state blockade mechanism in Sec.~\ref{two}.
Then we investigate the robustness for preparation of the maximally entangled state based on the ground-state Rybderg blockade in a closed system and in an open system, respectively in Secs.~\ref{three} and \ref{four}. And then, we directly generalize the above schemes to the case of three-atom entanglement in Sec.~\ref{five}. Finally, we give a summary of our proposal in Sec.~\ref{six}.

\section{Ground-state blockade mechanism between two atoms }\label{two}
\begin{figure}
\scalebox{0.48}{\includegraphics{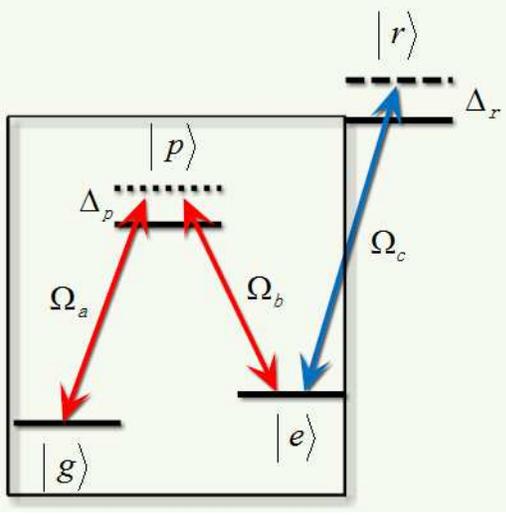} }
\caption{\label{p1}(Color online) Schematic view of atomic-level configuration. The ground states $|g\rangle$ and $|e\rangle$ are dispersively coupled to the excited state $|p\rangle$ with Rabi frequencies $\Omega_a$ and $\Omega_b$, respectively. An additional classical field drives the transition $|e\rangle\leftrightarrow|r\rangle$ with the Rabi
frequency $\Omega_{\rm c}$. $\Delta_{p(r)}$ represents the corresponding single-photon detuning parameter.}
\end{figure}
\begin{figure}
\scalebox{0.5}{\includegraphics{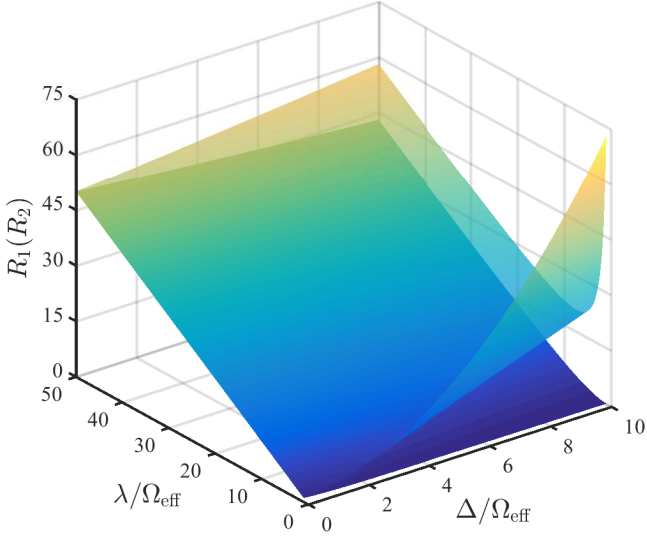} }
\caption{\label{p2}(Color online) The ratios $R_1$ (upper surface) and $R_2$ (lower surface) are plotted as functions
$\Delta/\Omega_{\rm eff}$ and $\lambda/\Omega_{\rm eff}$,
where ${R_{1}}=|E_{+}/(\sqrt{2}\Omega_{\rm eff}\sin\alpha)|$ and
${R_{2}}=|E_{-}/(\sqrt{2}\Omega_{\rm eff}\cos\alpha)|$.}
\end{figure}

We consider a system consisting of two $N$-type four level Rydberg atoms, and the relevant configuration
of atomic level is illustrated in Fig.~\ref{p1}.  The ground states $|g\rangle$ and $|e\rangle$ are dispersively coupled to the excited state $|p\rangle$  by two classical fields with Rabi frequencies $\Omega_a$, $\Omega_b$, and a common detuning $-\Delta_p$. And the ground state $|e\rangle$ can be pumped into the excited Rydberg state $|r\rangle$ by a driving field with Rabi
frequency $\Omega_{\rm c}$, detuned by $-\Delta_{r}$. In the interaction picture with respect to a rotating frame, the Hamiltonian of the system reads ($\hbar=1$)
\begin{eqnarray}\label{two01}
H_{I}&=&\sum_{i=1}^2 \Omega_a|p\rangle_{i}\langle g|+\Omega_b|p\rangle_{i}\langle e|+\Omega_c|r\rangle_{i}\langle e|+{\rm{H.c.}} \cr &&-\sum_{i=1}^2\Delta_p|p\rangle_{i}\langle p|+(U-2\Delta_r)|rr\rangle\langle rr|,
\end{eqnarray}
where $U$ represents the Rydberg-mediated interaction as two atoms simultaneously occupy the Rydberg state. This kind of nonlinear interaction originates from the dipole-dipole
potential with energy $C_{3}/r^{3}$ or the long-range vander Waals interaction $C_{6}/r^{6}$, with $r$ being the distance between two Rydberg
atoms, and $C_{3(6)}$ depending on the quantum numbers of the Rydberg state \cite{comparat2010dipole,PhysRevLett.110.263201}.
Through the standard second-order perturbation theory, we may
adiabatically eliminate the excited state $|p\rangle$ and the single-atom Rydberg state $|r\rangle$ in the regime of large detuning limit $\Delta_{p}\gg\{\Omega_{a},\Omega_{b}\}$, and $\Delta_{r}\gg\Omega_{c}$. Then we obtain an effective Hamiltonian as
\begin{eqnarray}\label{two02}
H_{\rm eff}&=&\sum_{i=1}^2\frac{\Omega_a^2}{\Delta_p}|g\rangle_{i}\langle g|+\bigg(\frac{\Omega_b^2}{\Delta_p}
+\frac{\Omega_c^2}{\Delta_r}\bigg)|e\rangle_{i}\langle e| \cr
&&+\bigg[\frac{2\Omega_c^2}{\Delta_r}|ee\rangle\langle rr|+\sum_{i=1}^2\frac{\Omega_a\Omega_b}
{\Delta_p}|g\rangle_{i}\langle e|+{\rm H.c.}\bigg] \cr
&&+\bigg(U-2\Delta_r+\frac{2\Omega_c^2}{\Delta_r}\bigg)|rr\rangle\langle rr|.
\end{eqnarray}
The first two terms will cause unwanted shifts to our system, which need to be canceled via introducing other ancillary levels. And the Stark shift in the last term $2\Omega_c^2/\Delta_r$ stems from the two-photon transition $|ee\rangle\leftrightarrow\langle rr|$. Now the above Hamiltonian can be rewritten in a concise form
\begin{equation}\label{two03}
H_{\rm eff}=\sum_{i=1}^2\Omega_{\rm eff}|g\rangle_{i}\langle e|+\lambda|ee\rangle\langle rr|+{\rm H.c.}+\Delta|rr\rangle\langle rr|,
\end{equation}
where $\Omega_{\rm eff}=\Omega_{a}\Omega_{b}/\Delta_{p}$, $\lambda={2\Omega_{c}^2/\Delta_{r}}$ and
$\Delta=U-2\Delta_{r}+{2\Omega_{c}^2/\Delta_{r}}$. We now divide Eq.~(\ref{two03}) into two parts, i.e.
$H_{\rm eff}=H_{\alpha}+H_{\beta}$, where
$H_{\alpha}=\sum_{i=1}^2\Omega_{\rm eff}(|g\rangle_{i}\langle e|+|e\rangle_{i}\langle g|)$ describes the Raman transition of two ground states and
$H_{\beta}=\lambda(|ee\rangle\langle rr|+|rr\rangle\langle ee|)+\Delta|rr\rangle\langle rr|$ represents the Rydberg antiblockade interaction. The Hamiltonian $H_{\beta}$ can be diagonalized by
the eigenstates $|\Psi_+\rangle=\cos\alpha|rr\rangle+\sin\alpha|ee\rangle$ and $|\Psi_-\rangle=\sin\alpha|rr\rangle-\cos\alpha|ee\rangle$,
corresponding to eigenvalues $E_+=(\Delta+\sqrt{\Delta^2+4\lambda^2})/2$ and $E_-=(\Delta-\sqrt{\Delta^2+4\lambda^2})/2$, respectively, and $\alpha=\arctan[2\lambda/(\Delta+\sqrt{\Delta^2+4\lambda^2})]$. Thus we have
\begin{eqnarray}\label{two04}
H_{\rm eff}&=&\sqrt{2}\Omega_{\rm eff}|gg\rangle\frac{1}{\sqrt{2}}(\langle ge|+\langle eg|)+\sqrt{2}\Omega_{\rm eff}(\sin\alpha|\Psi_+\rangle \cr
&&-\cos\alpha|\Psi_-\rangle) \frac{1}{\sqrt{2}}(\langle ge|+\langle eg|)+{\rm H.c.} \cr
&&+E_+|\Psi_+\rangle\langle \Psi_+|+E_-|\Psi_-\rangle\langle \Psi_-|.
\end{eqnarray}
It is shown that the ground state $|gg\rangle$ resonantly interacts with the  entangled state $(|ge\rangle+|eg\rangle)/\sqrt{2}$ with coupling constant $\sqrt{2}\Omega_{\rm eff}$, and $(|ge\rangle+|eg\rangle)/\sqrt{2}$ is then coupled to the state $|\Psi_+\rangle$ $(|\Psi_-\rangle )$ with strength $\sqrt{2}\Omega_{\rm eff}\sin\alpha$ $(\sqrt{2}\Omega_{\rm eff}\cos\alpha)$, detuning $E_+$ $(E_-)$. In the limits of $R_1=|E_+/(\sqrt{2}\Omega_{\rm eff}\sin\alpha)|\gg1$ and $R_2=|E_-/(\sqrt{2}\Omega_{\rm eff}\cos\alpha)|\gg1$, the high-frequency oscillating terms may be neglected and an approximated ground-state blockade Hamiltonian is obtained
\begin{eqnarray}\label{two06}
H_{gb}=\sqrt{2}\Omega_{\rm eff}|gg\rangle\frac{1}{\sqrt{2}}(\langle ge|+\langle eg|)+{\rm H.c.}.
\end{eqnarray}
In Fig.~\ref{p2}, the ratio $R_1$ $(R_2)$ is plotted as a function of
$\Delta/\Omega_{\rm eff}$ and $\lambda/\Omega_{\rm eff}$, which is explicit to determine the values of $\lambda$ and $\Delta$ so as to get a better ground-state blockade effect. For instance, Tab.~\ref{table1} lists the maximal populations of states $|T\rangle=(|ge\rangle+|eg\rangle)/\sqrt{2}$ and $|ee\rangle$ from the initial state $|gg\rangle$. The corresponding results are extracted from the numerical simulation of Eq.~(\ref{two01}), which signifies that $R_1$ $(R_2)=20$ is big enough for occurrence of ground-state blockade. In the following, we will reveal the advantage of ground-state blockade on preparation of quantum entanglement by setting $\Delta=0$ for simplicity.

\begin{table}
\centering
\scalebox{1.1}{
\begin{tabular}{p{1.2cm}p{1.2cm}p{1.2cm}p{1.2cm}l}
\hline\hline
$R_1$ &$R_2$ & $|gg\rangle$ & $|T\rangle$& $|ee\rangle$\\ \hline
10 &10 &1.00 &0.9673&0.0015 \\
20 &20 &1.00 &0.9916&$1.0121\times10^{-4}$\\
50 &50 &1.00 &0.9963&$2.7228\times10^{-6}$\\ \hline\hline
\end{tabular}}
\caption{\label{table1} Maximal populations of relevant quantum states corresponding three specific ratios of $R_1$ $(R_2)$ at $\Delta=0$.}
\end{table}
\section{Robust entanglement via shortcut to adiabatic passage}\label{three}
\begin{figure}
\scalebox{0.48}{\includegraphics{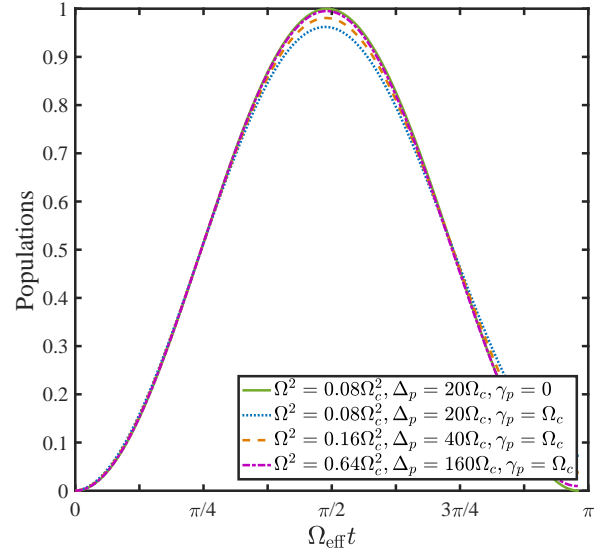} }
\caption{\label{p3}(Color online) Population of state $|e\rangle$ in the process of quantum state transfer for a single $\Lambda$-type atom versus a common dimensionless time $\Omega_{\rm eff}t$ with different detuning and decoherence parameters, where $\Omega=\Omega_a=\Omega_b$ is assumed for simplicity.}
\end{figure}
Before  preparation of entangled state, let us first discuss the robustness of quantum state transfer for a single $\Lambda$-type atom, in the presence of spontaneous emission. It has a guiding significance on the choice of parameters for experimental realization of entanglement.
The studied system has been shown in the box of Fig.~\ref{two01}, the atom can spontaneously decay  with the same rate $\gamma_p/2$ from excited
state $|p\rangle$ into the ground states $|g\rangle$ and $|e\rangle$, respectively. Hence the complete master equation describing the dynamics of this system reads
\begin{eqnarray}\label{threeA02}
\dot{\rho} _{sgl}&=&-i[H_{sgl},{\rho} _{sgl}]+\frac{\gamma_p}{2}\sum_{m=g,e}{\cal D}[\sigma_-^{m}]\rho_{sgl},
\end{eqnarray}
where
$
H_{sgl}=\Omega_{a}|p\rangle\langle g|+\Omega_{b}|p\rangle\langle e|+{\rm{H.c.}}-\Delta_{p}|p\rangle\langle p|,
$
and ${\cal D}[\sigma_-^{m}]\rho_{sgl}=\sigma_-^{m}\rho_{sgl}\sigma_+^{m}-\{\sigma_+^{m}\sigma_-^{m}, \rho_{sgl}\}/2$.
$\sigma_-^{m}=(\sigma_+^{m})^{\dag}$ is the lowering operator of atom from the excited state $|p\rangle$ to the ground state $|m\rangle$. After adiabatically eliminating the excited state $|p\rangle$ under the large detuning condition $\Delta_p\gg\{\Omega_a,\Omega_b\}$, the single-atom master equation is reduced to
\begin{eqnarray}\label{mas}
\dot{\rho}_{sgl}&=&-i[H_{rd},{\rho_{sgl}}]+\sum_{m=g,e}{\cal D}[R_{mp}]\rho_{sgl},
\end{eqnarray}
where $H_{rd}$ denotes the effective Hamiltonian of Raman transition between  states $|g\rangle$ and $|e\rangle$ with coupling strength $\Omega_{\rm eff}$, and
\begin{equation}\label{decay}
R_{mp}=\sqrt{\frac{\gamma_p}{2}}|m\rangle\bigg(\frac{\Omega_a}{\Delta_p}\langle g|+\frac{\Omega_b}{\Delta_p}\langle e|\bigg),\  (m=g,e)
\end{equation}
represents the effective decay operator \cite{PhysRevA.76.022331,stevenson2011production,PhysRevA.94.032307}. Eq.~(\ref{decay}) gives a quantitative relationship among the Rabi frequency of classical fields, the frequency detuning parameter, and the spontaneous emission rate of atom. It can be directly seen that the decaying rate is reduced to
\begin{equation}\label{efdecay}
\gamma_{\rm eff}=\frac{\gamma_p}{2}\times\frac{\Omega^2}{\Delta^2_p}=\frac{\gamma_p}{2\Delta_p}\times\Omega_{\rm eff},
\end{equation}
 where we have assumed $\Omega=\Omega_{a(b)}$ for the sake of convenience. Therefore, we may reduce the effect of spontaneous emission by enlarging the value of detuning $\Delta_p$ for implementing the quantum state transfer, even without changing the interaction time of system. Fig.~\ref{p3} characterizes the population $P_e(t)=\langle e|\rho_{sgl}(t)|e\rangle$ of state $|e\rangle$ in the process of quantum state transfer from the initial state $|g\rangle$ corresponding to different detuning and decoherence parameters The effective Raman coupling strength is fixed at $\Omega_{\rm eff}=0.004\Omega_c$. For $\Delta_p=20\Omega_c$, the maximal state transfer efficiency is 96.21\% as $\gamma_p=\Omega_c$ (dotted line), and this value is promoted to 99.51\% for $\Delta_p=160\Omega_c$ (dash-dotted line), which is very close to the ideal case 99.96\% (solid line). Hence one can see that, a large $\Delta_p$ does provide an immune way to the spontaneous emission of atom.

\begin{figure}
\scalebox{0.48}{\includegraphics{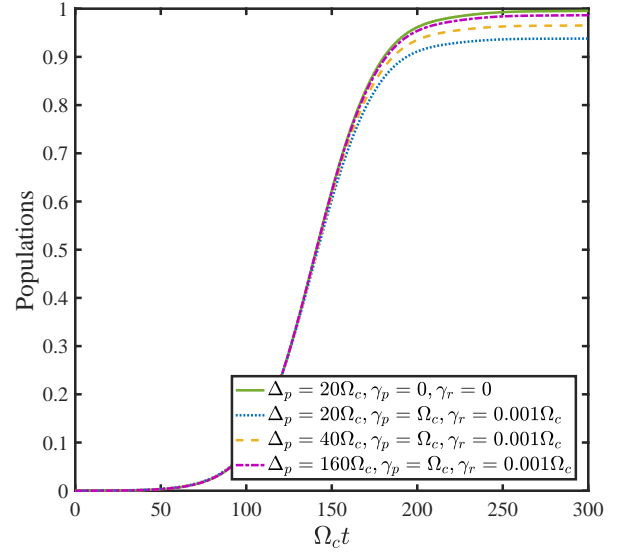} }
\caption{\label{p4}(Color online) Population of the maximally entangled state $(|ge\rangle+|eg\rangle)/\sqrt{2}$ during the shortcut to adiabatic passage versus dimensionless interaction time $\Omega_{\rm c}t$ for different detuning and decoherence parameters, and we have chosen $t_{\rm c}=300/\Omega_c$, $\tau=0.2t_{\rm c}$, $T=0.3t_{\rm c}$ in Eqs.~(\ref{driv1}) and (\ref{driv2}) and $\Delta_r=20\Omega_c$.}
\end{figure}

The technology of shortcut to adiabatic passage permits a fast manipulation of quantum states in a robust way against the fluctuation of parameters \cite{PhysRevLett.105.123003,PhysRevA.89.033856,PhysRevA.89.012326,du2016experimental,PhysRevA.93.052109}. In order to design a counteradiabatic Hamiltonian that can be realized in experiment,
we first consider a toy model below
\begin{equation}\label{threeB01}
H_{ap}(t)=\sqrt{2}{\Omega_a^{'}(t)}|gg\rangle\langle \Phi|+\Omega_b^{'}(t)|T\rangle\langle \Phi|
+{\rm H.c.},
\end{equation}
where $|\Phi\rangle=(|gp\rangle+|pg\rangle)/\sqrt{2}$. This Hamiltonian is equivalent to a simple three-level
system with an excited state $|\Phi\rangle$ and two ground states $|gg\rangle$ and $|T\rangle$.
The corresponding eigenstates can be easily obtained
\begin{eqnarray}\label{threeB02}
|n_0(t)\rangle&=&\cos[\theta(t)]|gg\rangle-\sin[\theta(t)]|T\rangle,\\
|n_{\pm}(t)\rangle&=&\frac{\sin[\theta(t)]}{\sqrt{2}}|gg\rangle\pm\frac{1}{\sqrt{2}}|\Phi\rangle+\frac{\cos[\theta(t)]}{\sqrt{2}}
|T\rangle,
\end{eqnarray}
and the eigenvalues are
$\varepsilon_0=0$, $\varepsilon_{\pm}=\pm\Omega^{'}$, respectively, where $\theta(t)=\arctan[\sqrt{2}\Omega_a^{'}(t)/\Omega_b^{'}(t)]$ and $\Omega^{'}=\sqrt{2\Omega_a^{'2}(t)+\Omega_b^{'2}(t)}$. According to Berry's transitionless tracking algorithm \cite{berry2009transitionless}, the simplest form of reverse engineering Hamiltonian $H_{cap}(t)$,
 which is related to the original Hamiltonian $H_{ap}(t)$, takes the form
\begin{eqnarray}\label{threeB03}
H_{cap}(t)&=&i\sum_{k=0,\pm}|\partial_tn_k(t)\rangle\langle n_k(t)|\nonumber\\
&=&i\dot{\theta}(t)|gg\rangle\frac{1}{\sqrt{2}}(\langle ge|+\langle eg|)+{\rm H.c.},
\end{eqnarray}
where $\dot{\theta}(t)=\sqrt{2}[\dot{\Omega}_a^{'}(t)\Omega_b^{'}(t)-\Omega_a^{'}(t)\dot{\Omega}_b^{'}(t)]/\Omega^{'2}$. Comparing Eq.~(\ref{threeB03}) with Eq.~(\ref{two06}), we are able to obtain
an alternative physically feasible Hamiltonian whose effect is equivalent to $H_{\rm cap}(t)$
\begin{eqnarray}\label{threeB05}
\widetilde{H}_{\rm eff}= i\frac{\Omega_{cap}^{2}}{\Delta_{\rm p}}|gg\rangle\frac{1}{\sqrt{2}}(\langle ge|+\langle eg|)+ {\rm H.c.},
\end{eqnarray}
and the shortcut to adiabatic passage for preparation of bipartite entanglement
can be achieved as long as $\Omega_a=i\Omega_{cap}/\sqrt{2}$, $\Omega_b=\Omega_{cap}$, and ${\Omega_{cap}^{2}}/{\Delta_{p}}=\dot{\theta}(t)$, i.e.
\begin{eqnarray}\label{threeB06}
\Omega^2_{cap}={{\Delta_{\rm p}}\dot{\theta}(t)}={\frac{\sqrt{2}\Delta_{\rm p}[\dot{\Omega}_a^{'}(t)\Omega_b^{'}(t)-\Omega_a^{'}(t)\dot{\Omega}_b^{'}(t)]}{\Omega^{'2}}},
\end{eqnarray}
where the Rabi frequencies $\Omega_{a}^{'}(t)$ and
$\Omega_{b}^{'}(t)$ are chosen as
\begin{equation}\label{driv1}
\Omega_{a}^{'}(t)=\Omega_0\exp\bigg[-\frac{(t-t_c/2-\tau)^2}{T^2}\bigg],
\end{equation}
\begin{equation}\label{driv2}
\Omega_{b}^{'}(t)=\Omega_0\exp\bigg[-\frac{(t-t_c/2+\tau)^2}{T^2}\bigg],
\end{equation}
in order to satisfy the boundary condition of the stimulated Raman adiabatic passage on the one hand, and meet the requirement of the following ground-state blockade effect for time-dependent Raman couplings on the other hand \cite{PhysRevA.75.032302},
\begin{equation}\label{detund}
\bigg|\frac{1}{2}S(t_c)\bigg|=\bigg|\frac{i}{2}{\int_0^{t_c}e^{-i\lambda(t_c-t)}}\frac{\Omega_a(t)\Omega_b(t)}{\Delta_p}dt\bigg|\ll1.
\end{equation}
We remark that Eq.~(\ref{detund}) automatically degenerates to $|\Omega_{\rm eff}|\ll|\lambda|$ for the time-independent Raman couplings of Eq.~(\ref{two06}) in the absence of $\Delta$.
In Fig.~\ref{p4}, we check the performance of the shortcut to adiabatic passage in generation of entangled state $(|ge\rangle+|eg\rangle)/\sqrt{2}$ from the initial state $|gg\rangle$ by setting the operation time $t_c=300/\Omega_c$, $\tau=0.2t_{\rm c}$, and $T=0.3t_{\rm c}$.
With the dissipation being considered, a conclusion as the same as the single-atom case can be made that a large detuning condition guarantees a high fidelity $F(t)={\rm Tr}\sqrt{\rho^{1/2}\rho(t)\rho^{1/2}}=\sqrt{P(t)}=99.32\%$, corresponding to the dash-dotted line.

\section{steady entanglement via quantum-jump-based feedback control}\label{four}
\begin{figure}
\scalebox{0.48}{\includegraphics{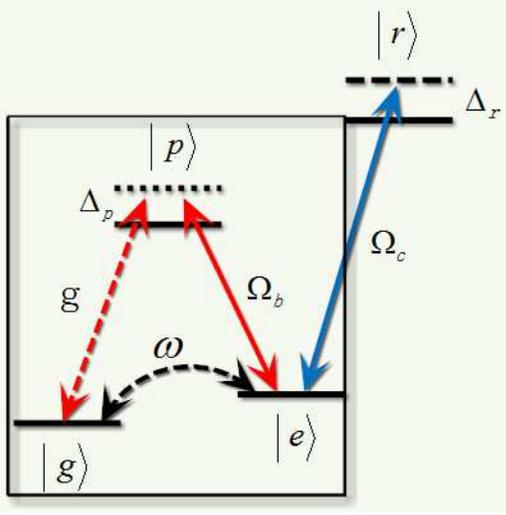} }
\caption{\label{p5}(Color online) Schematic view of the atomic-level configuration.
Compared with Fig.~\ref{p1},
the transition $|g\rangle\leftrightarrow|p\rangle$ is repalced by a quantized cavity field mode with coupling strength $g$, and a resonant transition between $|g\rangle$ and $|e\rangle$
is driven by a microwave field with Rabi frequency $\omega$.}
\end{figure}
The above analysis has demonstrated that a regime of ground-state blockade effect functioning well is also immune to the atomic decay. Therefore
combined with cavity quantum electrodynamics, the ground-state blockade will provides a novel approach to quantum state preparation, especially for the cavity-loss-induced generation of entangled atoms \cite{Carvalho2007,Carvalho2008,Stevenson2011}. In this section, we consider an atom-cavity interaction system, as depicted in Fig.~{\ref{p5}}.
 The transition between
the levels $|g\rangle\leftrightarrow|p\rangle$ is coupled to the cavity mode resonantly with coupling constant $g$. The transition
$|e\rangle\leftrightarrow|p\rangle$ and $|e\rangle\leftrightarrow|r\rangle$ are driven by a nonresonant classical laser field with Rabi
frequencies $\Omega_{b}$ and $\Omega_{c}$, respectively. The resonant coupling between ground states $|g\rangle$ and $|e\rangle$
is realized by a microwave field with Rabi frequency $\omega$. Thus the master equation of
system could be written as
\begin{eqnarray}\label{four04}
\dot{\rho}&=&-i[H_I^{'},\rho]+\frac{\gamma_p}{2}\sum_{n=1}^2\big({\cal D}[|g\rangle_n\langle p|]\rho+{\cal D}[|e\rangle_n\langle p|]\rho\big)\nonumber\\&&+\gamma_r\sum_{n=1}^2{\cal D}[|e\rangle_n\langle r|]\rho+\kappa{\cal D}[a]\rho,
\end{eqnarray}
where the Hamiltonian
$
H_{I}^{'}=\sum_{i=1}^2 g|p\rangle_{i}\langle g|a+\Omega_b|p\rangle_{i}\langle e|+\Omega_c|r\rangle_{i}\langle e|
+\omega|g\rangle_{i}\langle e|+{\rm{H.c.}}-\Delta_p|p\rangle_{i}\langle p|+(U
-2\Delta_r)|rr\rangle\langle rr|
$, $\gamma_r$ is the decaying rate of the Rydberg state, $a$ is the annihilation operator of cavity mode, and $\kappa$ is the loss rate of cavity.
After adiabatically eliminating the excited state $|p\rangle$ and the single-atom state  $|r\rangle$,
we have
\begin{eqnarray}\label{four02}
H_{\rm eff}^{'}&=&\sum_{i=1}^2(g_{\rm eff}a^{\dag}+\omega)|g\rangle_{i}\langle e|+\lambda|ee\rangle\langle rr|+{\rm H.c.},
\end{eqnarray}
where $g_{\rm eff}={g\Omega_{b}}/{\Delta_{p}}$. In the regime of ground-state blockade, $\{g_{\rm eff},\omega\}\ll\lambda$,
the double occupation of state $|ee\rangle$ is suppressed and the above Hamiltonian is further simplified to
\begin{eqnarray}\label{four03}
H_{gb}^{'}=(g_{\rm eff}a^{\dag}+\omega)|gg\rangle(\langle ge|+\langle eg|)+{\rm H.c.}.
\end{eqnarray}
In this case, the effective master equation prompting the evolution of two atoms becomes
\begin{eqnarray}\label{four04}
\dot{\rho}_r&=&-i[H_{gb}^{'},\rho_r]+\sum_{n=1}^2{\cal D}[R_{gp}^{'n}]\rho_r+{\cal D}[R_{ep}^{'n}]\rho_r\nonumber\\
&&+\kappa{\cal D}[a]\rho_r,
\end{eqnarray}
\begin{figure}
\scalebox{0.48}{\includegraphics{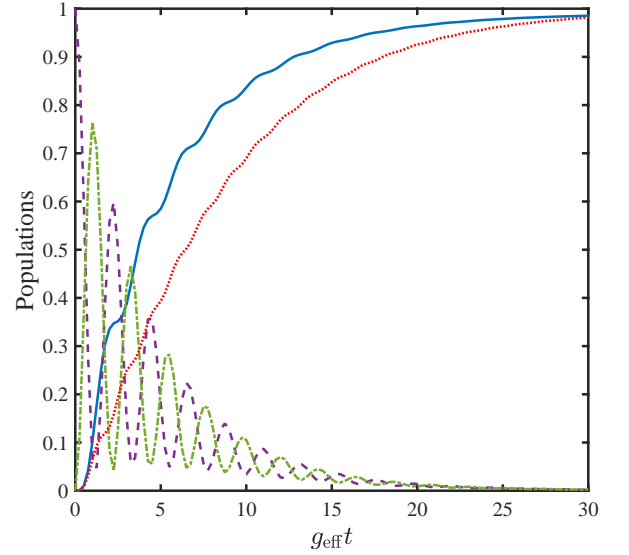}}
\caption{\label{p6}(Color online) Populations of quantum states versus dimensionless time $g_{\rm eff}t$ during preparation of the
antisymmetric entangled state $|S\rangle$. Other parameters: $\omega=g_{\rm eff}$, $\kappa=\lambda=10g_{\rm eff}$ and $\eta=-0.5\pi$.}
\end{figure}

\begin{figure*}
\begin{minipage}[t]{0.49\linewidth}
\centering
\includegraphics[scale=0.48]{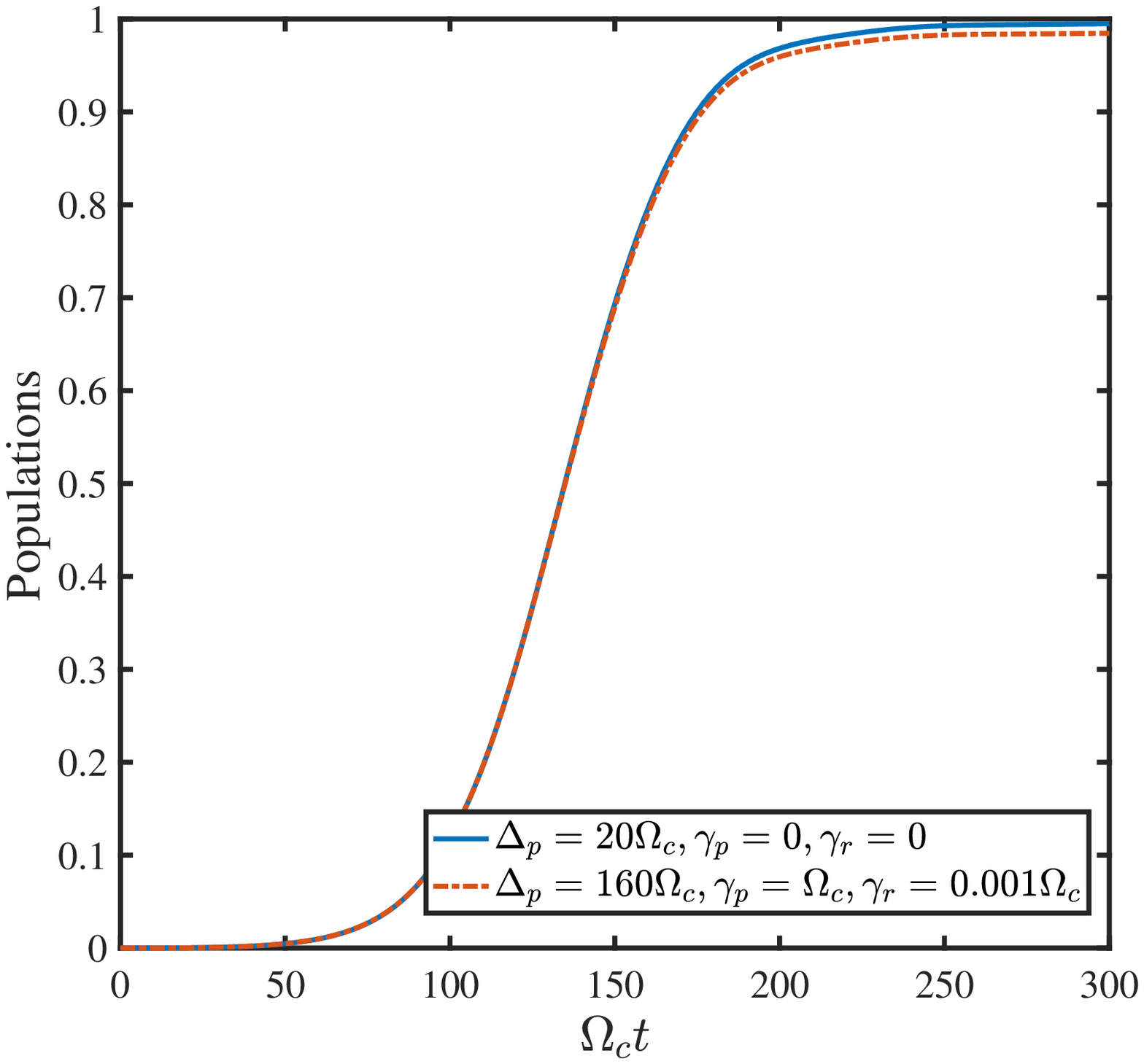}
 \centerline{(a)}
\end{minipage}
\begin{minipage}[t]{0.49\linewidth}
\centering
\includegraphics[scale=0.48]{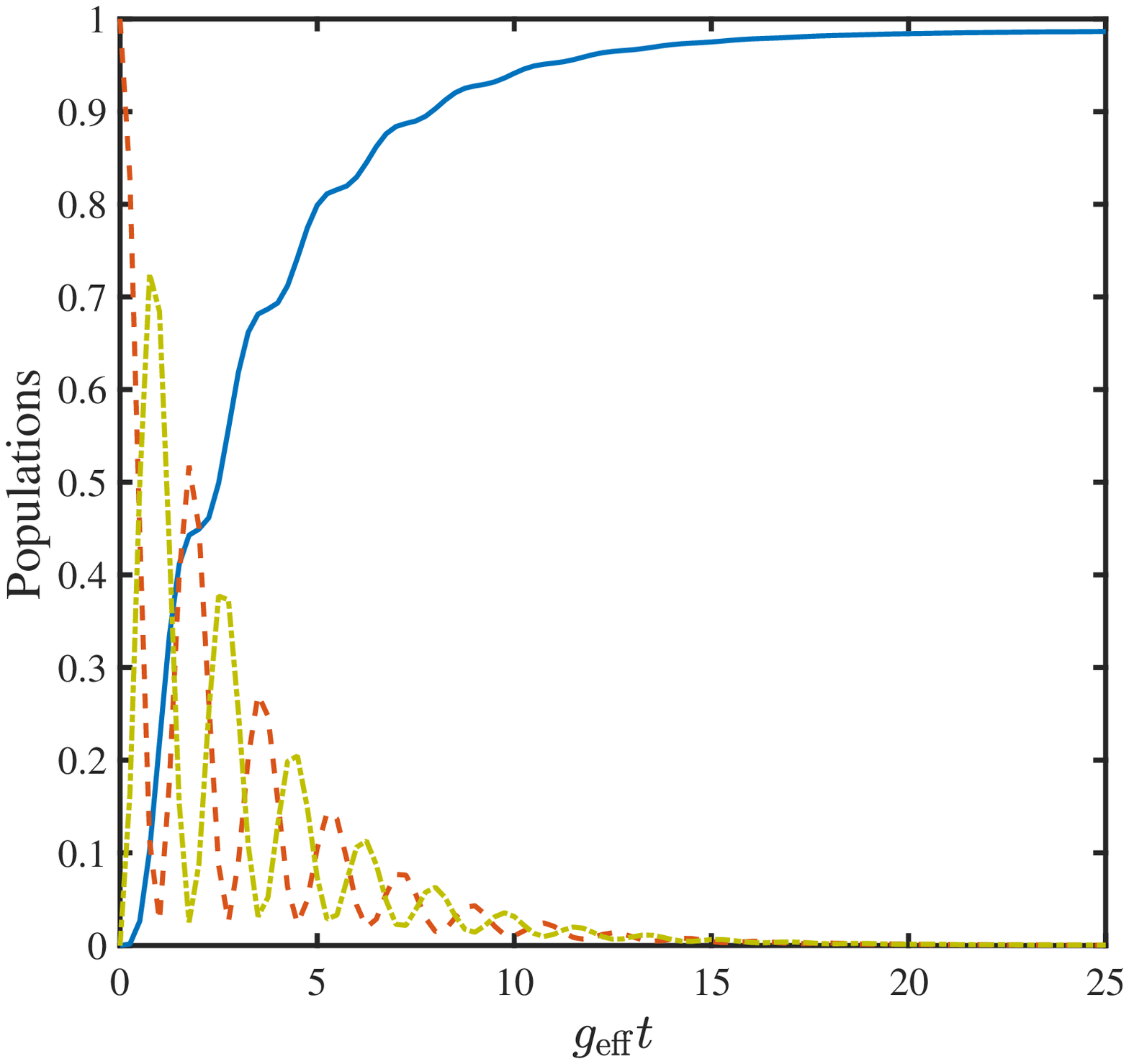}
 \centerline{(b)}
\end{minipage}
\caption{\label{p7}(a) Population of $W$ state during the shortcut to adiabatic passage versus dimensionless interaction time $\Omega_{\rm c}t$. Other parameters
are the same as those in Fig.~\ref{p4}. (b) Populations of quantum states versus dimensionless time $g_{\rm eff}t$ during preparation of the three-atom
decoherence-free state $|\rm DFS\rangle$. Other parameters: $\omega=g_{\rm eff}$, $\kappa=10g_{\rm eff}$, $\lambda=20g_{\rm eff}$, and $\eta=-0.5\pi$.}
\end{figure*}
with
\begin{eqnarray}\label{four05}
R_{gp}^{'}=\frac{\Omega_b}{\Delta_p}\sqrt{\frac{\gamma_p}{2}}|g\rangle\langle e|,~~~~
R_{ep}^{'}=\frac{\Omega_b}{\Delta_p}\sqrt{\frac{\gamma_p}{2}}|e\rangle\langle e|,
\end{eqnarray}
being the effective decay operators from $|e\rangle$ to $|g\rangle$, and $|e\rangle$ to $|e\rangle$, respectively. For a
strongly damped cavity mode, $\kappa\gg \{g_{\rm eff}, \omega\}$,
we further adiabatically eliminate the populations of cavity mode, and acquire the master
equation for the reduced density operator of atoms
\begin{equation}\label{four06}
\dot{\rho}_r=-i\omega[(J_++J_-),\rho_r]+\Gamma{\cal D}[J_-]\rho_r,
\end{equation}
where $J_-=J_+^{\dag}=|gg\rangle(\langle eg|+\langle ge|)$ is the collective
lowing operators of atom, and $\Gamma=4g^2_{\rm eff}/\kappa$ is the collective amplitude damping rate. In Eq.~(\ref{four06}), we also neglect the spontaneous emission terms by supposing $\Gamma\gg\gamma_p\Omega_b^2/(2\Delta_p^2)$
Once the local feedback scenario is introduced, the cavity output will be measured by a
photodetector whose signal provides the input to the application
of the feedback operator $U_{\rm fb}=\exp[-i\eta(\sigma_x\otimes I)]$, and the unconditioned master equation for
this case is derived
\begin{equation}\label{four07}
\dot{\rho}_r=-i\omega[(J_++J_-),\rho_r]+\Gamma{\cal D}[U_{\rm fb}J_-]\rho_r.
\end{equation}
Note the local feedback operator is
approximated to
$
U_{\rm fb}=\exp[-i\eta(|g\rangle_{1}\langle e|+|e\rangle_{1}\langle g|)\otimes|g\rangle_{2}\langle g|],
$
because of the ground-state blockade effect. A simple inspection shows $|S\rangle=(|ge\rangle-|eg\rangle)/\sqrt{2}$ is the unique stationary state solution
of Eq.~(\ref{four07}). In Fig.~\ref{p6}, we numerically simulate the populations of quantum states versus time $g_{\rm eff}t$ during the preparation of the
antisymmetric entangled state $|S\rangle$ from a initial state $|gg\rangle$ with parameters $\omega=g_{\rm eff}$, $\kappa=\lambda=10g_{\rm eff}$ and $\eta=-0.5\pi$. It only takes $t=13/g_{\rm eff}$ to make the population of state $|S\rangle$ exceed 90\% for the current scheme (solid line), compared with $t=18/g_{\rm eff}$ for the case without considering ground-state blockade (dotted line). In this sense, the effect of ground-state blockade can speed up the convergence time for state preparation in an open system.

\section{generalization to three-atom entanglement}\label{five}
In the scheme of utilizing shortcut to adiabatic passage, an three-atom $W$ state $(|egg\rangle+|geg\rangle+|gge\rangle)/\sqrt{3}$ can be prepared straightforwardly with the following time-dependent Hamiltonian
\begin{eqnarray}\label{threeB05}
\widetilde{H}_{\rm eff}= \frac{\sqrt{3}\Omega_{a}(t)\Omega_b(t)}{\Delta_{\rm p}}|ggg\rangle\langle W|+ {\rm H.c.}.
\end{eqnarray}
The counteradiabatic Hamiltonian is then received by selecting $\Omega_a=i\Omega_{cap}/\sqrt{3}$, $\Omega_b=\Omega_{cap}$, and
\begin{eqnarray}\label{threeB06}
\Omega^2_{cap}={{\Delta_{\rm p}}\dot{\theta}(t)}={\frac{\sqrt{3}\Delta_{\rm p}[\dot{\Omega}_a^{'}(t)\Omega_b^{'}(t)-\Omega_a^{'}(t)\dot{\Omega}_b^{'}(t)]}{\Omega^{'2}}},
\end{eqnarray}
where $\Omega^{'}=\sqrt{3\Omega_a^{'2}(t)+\Omega_b^{'2}(t)}$.
At the same time, the condition of ground-state blockade should be satisfied
\begin{equation}\label{detun}
\bigg|\frac{i}{2}{\int_0^{t_c}e^{-i\lambda(t_c-t)}}\frac{\sqrt{2}\Omega_{a}(t)\Omega_{b}(t)}{\Delta_p}dt\bigg|\ll1.
\end{equation}

As for the quantum-feedback-based scheme, the local feedback operator on the first atom $
U_{\rm fb}=\exp[-i\eta(|g\rangle_{1}\langle e|+|e\rangle_{1}\langle g|)\otimes|g_2\rangle\langle g_2|\otimes|g_3\rangle\langle g_3|],
$ along with the dissipation of cavity will stabilize the system into a dark state of the collective lowing operator $J_-=$$|ggg\rangle$$(\langle egg|+\langle geg|+\langle gge|)$, i.e.
\begin{equation}\label{five02}
|{\rm DFS}\rangle_{N}=\frac{1}{\sqrt{6}}(|gge\rangle+|geg\rangle-2|egg\rangle).
\end{equation}

Fig.~\ref{p7} shows the population of three-atom entanglement as a function of time both for the closed system and the open system. On the left panel, the solid line indicates an ideal situation for the shortcut to adiabatic passage without dissipation, and the final fidelity of entangled state is 99.74\%.  Even in the presence of spontaneous emission $\gamma_p=\Omega_c$ and $\gamma_r=0.001\Omega_c$, a large detuning $\Delta_p=160\Omega_c$ preserves the fidelity up to 99.23\% (dash-dotted line). On the right panel, starting from the initial state $|ggg\rangle$, the population of state $|ggg\rangle$ (dashed line) and the $W$ state (dash-dotted line) undergo rapid coherent oscillation with an envelope decaying, while the three-atom decoherence-free state $|{\rm DFS}\rangle_3$ (solid line) converges to 99.32\% at a short time $t=25/g_{\rm eff}$ with parameters $\omega=g_{\rm eff}$, $\kappa=10g_{\rm eff}$, $\lambda=20g_{\rm eff}$ and $\eta=-0.5\pi$.

\begin{figure}
\scalebox{0.3}{\includegraphics{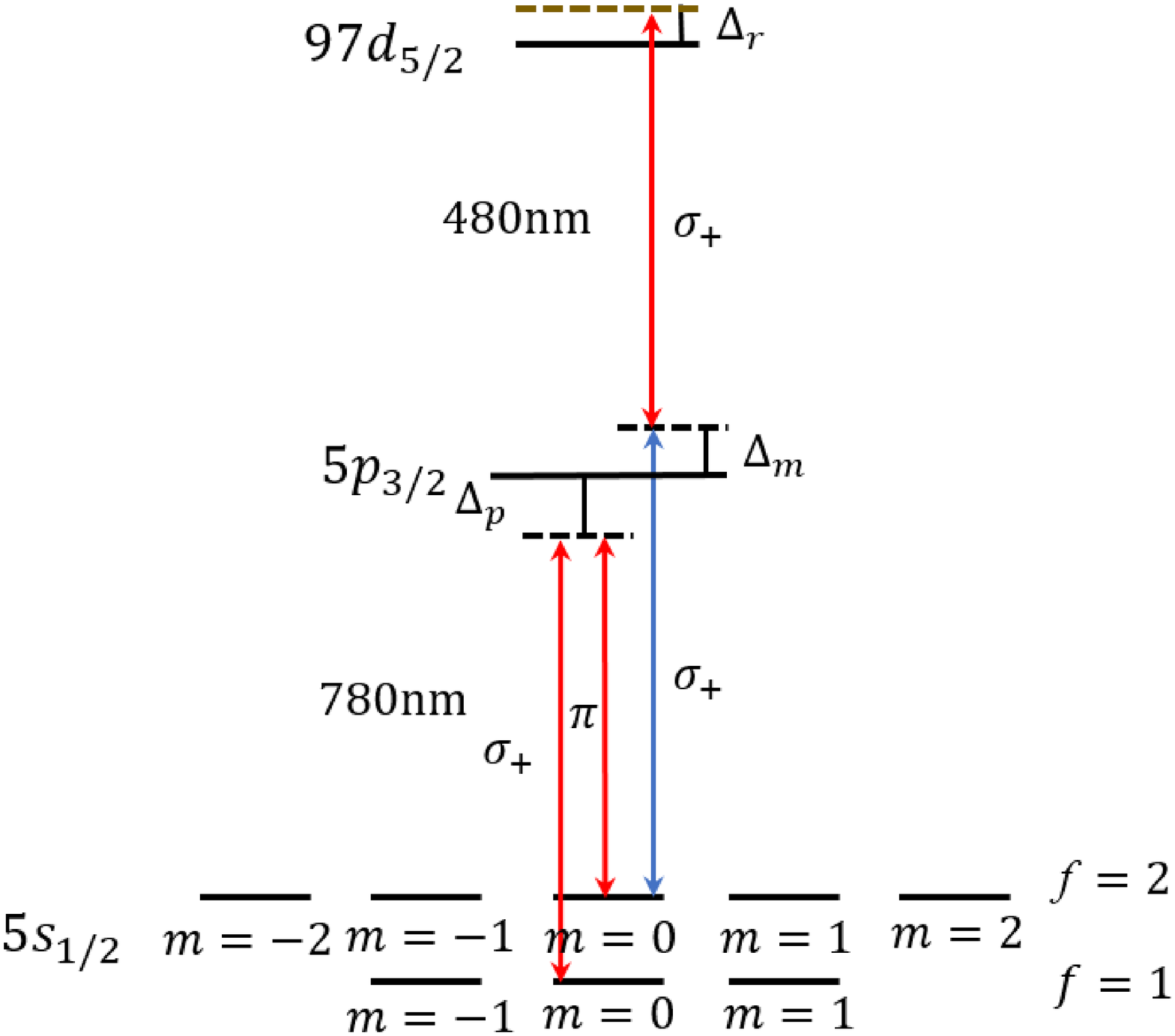} }
\caption{\label{p8}(Color online) Schematic view of the $N$-type Rydberg atom with relevant energy level structure of $^{87}$Rb atom.}
\end{figure}
In experiment, the configuration of $N$-type Rydberg atom can be found in $^{87}$Rb atom. The key components  of our proposal are the Raman transition of two ground states and a two-photon transition between ground state and rydberg state. In Ref.~\cite{PhysRevLett.96.063001}, the authors demonstrate a fast Rabi flopping at MHz between 5$s_{1/2}$ ground hyperfine states $|0\rangle=|f=1,m=0\rangle$ and $|1\rangle=|f=2,m=0\rangle$ that separated by $6.83$~GHz of neutral $^{87}$Rb atom, where each ground state is coupled to the 5$p_{3/2}$ excited by a detuning $\Delta_p=2\pi\times41$~GHz. In Refs.~\cite{gaetan2009observation,PhysRevLett.104.010502}, A. Browaeys {\it et al.} excite a ground state of 5$s_{1/2}$  to the Rydberg state of 58$d_{3/2}$ via a two-photon transition mediated by the optical state of 5$p_{1/2}$, where an effective two-photon Rabi frequency $\Omega_c\approx2\pi\times7$~MHz is achieved. In Refs.~\cite{PhysRevLett.104.010503,PhysRevA.82.030306}, the entanglement of two neutral atoms and corresponding  controlled-not gate are also demonstrated with $N$-type Rydberg atoms. Referring to our model, the relevant energy level structure is shown in Fig.~\ref{p8}, the ground states $|g\rangle$ and $|e\rangle$ correspond to atomic levels
$|f=1,m=0\rangle$, and $|f=2,m=0\rangle$ of 5$s_{1/2}$ manifold, the excited state $|p\rangle$ corresponds to 5$p_{3/2}$ atomic state with a radiative decaying rate $\gamma_p=2\pi\times3$~MHz, and the decaying rate of the 97$d_{5/2}$ Rydberg state $\gamma_r\sim2\pi\times1$~kHz. The Raman transition between ground states $|g\rangle$ and $|e\rangle$ is accomplished by $\sigma_+$ polarized and $\pi$ polarized 780nm laser beams both tuned to transit towards $|f=2,m=0\rangle$ of 5$p_{3/2}$ by about $\Delta_p=2\pi\times3.2$~GHz. The Rydberg excitation uses $\sigma_+$ polarized 780 and 480 nm
beams tuned for excitation of the Rydberg state of 97$d_{5/2}$, detuned by $\Delta_r=2\pi\times200$~MHz, leading to
the coupling strength between $|e\rangle$ and $|r\rangle$ of order $\Omega_c\sim2\pi\times10$~MHz. Note that a two-photon transition between the ground state $|g\rangle$ and the excited Rydberg state $|r\rangle$ cannot happen due to a large detuning parameter $\Delta_p+\Delta_m-\Delta_r$ on one hand, and the $\sigma_+$ polarized 480~nm laser beams is unable to  couple $|f=2,m=0\rangle$ of 5$p_{3/2}$ to other hyperfine levels of 97$d_{5/2}$ based on the selection rule on the other hand. For the first scheme governed by shortcut to adiabatic passage, the Rabi frequency $\Omega_{cap}$ is completely determined by the value of detuning parameter $\Delta_p$, provided the operation time $t_c$ is fixed. Hence we can obtain a high fidelity of two-atom entanglement $99.14\%$. For the second scheme based on quantum feedback control, the experimentally available coupling strength between atom and cavity $g=2\pi\times14.4$~MHz and the cavity decaying rate $\kappa=2\pi\times0.66$~MHz should also be taken into account \cite{brennecke2007cavity,PhysRevA.82.053832,PhysRevLett.110.090402}. In this case, we choose $\Delta_p=2\pi\times1.44$~GHz and $\Omega_b=g$ in order to gain a fidelity 98.95\% at a short time about $t=50/g_{\rm eff}\approx55.26~\mu$s.

We remark that the theoretical assumption $\Delta=(U-2\Delta_{r}+{2\Omega_{c}^2/\Delta_{r}})=0$ made throughout this paper is only for the sake of convenient discussion. In fact, the Rydberg-mediated interaction $U$ does not need to be limited to a specific value, as long as the approximation in Eq.~(\ref{two06}) is effective. We take $\Delta=5\Omega_{\rm eff}$ as an example, which can be extracted from Fig.~\ref{p2}. In this case, a selection of $\lambda=20\Omega_{\rm eff}$ corresponding $R_1\approx24.21$, $R_2\approx16.65$ is able to block the maximal population of state $|ee\rangle$ at 5.18$\times10^{-4}$. In this sense the mechanism of ground-state blockade proposed here can be implemented for a wide range of parameters.

\section{Summary}\label{six}
In summary, we have investigated how to actualize a ground-state blockade effect via a weak Raman transition and a strong Rydberg antiblockade.
This mechanism has prominent advantages in preparation of quantum entangled state, since it reserves the nonlinear Rydberg interaction and simultaneously provides a robust approach against the spontaneous emission of atom. In our future study, we will concentrate on the application of ground-state blockade in terms of quantum computing and quantum algorithm. We expect that our work may bring some new ideas on the quantum information processing with neutral atoms.

\section*{ACKNOWLEDGMENTS}
The authors thank the anonymous reviewer for constructive
comments that helped to improve the quality of this paper.
This work is supported by the Natural Science Foundation of China under Grants No.~11647308, No.~11674049,
No.~11534002, and No.~61475033, and by Fundamental Research Funds for the Central Universities under Grant No.
2412016KJ004.

\appendix*
\section{DETAIL DERIVATION OF THE GROUND-STATE BLOCKADE HAMILTONIAN}
In this appendix, we will give the detail derivation of the ground-state blockade Hamiltonian of Eq.~(\ref{two06}). According to Fig.~{\ref{p1}},
the Hamiltonian of our system in the Schr\"{o}dinger picture reads ($\hbar=1$)
\begin{eqnarray}\label{A1}
H_{S}&=&H_0+H_I,\\
H_{0}&=&\sum_{i=1}^2\omega_g|g\rangle_i\langle g|+\omega_e|e\rangle_i\langle e|+\omega_p|p\rangle_i\langle p|+\omega_r|r\rangle_i\langle r|,\nonumber\\
H_{I}&=&\sum_{i=1}^2 \Omega_a|p\rangle_{i}\langle g|e^{-i\omega_at}+\Omega_b|p\rangle_{i}\langle e|e^{-i\omega_bt}\nonumber\\&&+\Omega_c|r\rangle_{i}\langle e|e^{-i\omega_ct}+{\rm{H.c.}}+U|rr\rangle\langle rr|,\nonumber
\end{eqnarray}
where $\omega_{j}~(j=g,e,p,r)$ describes the frequency of atomic level $|j\rangle$ and $\omega_{k}~(k=a,b,c)$ represents the driving frequency of classical field corresponding to Rabi frequency $\Omega_k$. Thus in the interaction picture, we have
\begin{eqnarray}\label{A2}
H_{I}&=&H_I^0+H_I^1,\\
H_{I}^0&=&\sum_{i=1}^2 \Omega_a|p\rangle_{i}\langle g|e^{-i\Delta_pt}+\Omega_b|p\rangle_{i}\langle e|e^{-i\Delta_pt}+{\rm{H.c.}},\nonumber\\
H_{I}^1&=&\sqrt{2}\Omega_c|ee\rangle\langle \chi|e^{i\Delta_rt}+\sqrt{2}\Omega_c|rr\rangle\langle \chi|e^{i(\Delta_r+\delta)t}+{\rm{H.c.}},\nonumber
\end{eqnarray}
where we have introduced state $|\chi\rangle=(|er\rangle+|re\rangle)/{\sqrt{2}}$ for simplicity, and assumed the detuning parameters $\Delta_p=\omega_a-(\omega_p-\omega_g)=\omega_b-(\omega_p-\omega_e)$, $\Delta_r=\omega_c-(\omega_r-\omega_e)$, and $U=(2\Delta_r+\delta)$. Now the Hamiltonian $H_I$  has been divided into two parts, one part $H_{I}^0$ is the Raman interaction of atoms and the other part $H_{I}^1$ is the two-photon transition. In the regime of large detuning limit $\Delta_{p}\gg\{\Omega_{a},\Omega_{b}\}$, we can
adiabatically eliminate the excited state $|p\rangle$ and obtain the effective form of $H_{I}^0$ with Stark-shift term of state $|g\rangle~(|e\rangle)$, and effective Rabi frequency
\begin{eqnarray}\label{A3}
\frac{\langle g(e)|H_I^0|p\rangle\langle p|H_I^0|g(e)\rangle}{\Delta_p}&=&\frac{\Omega_{a(b)}^2}{\Delta_p},\\
\frac{\langle g(e)|H_I^0|p\rangle\langle p|H_I^0|e(g)\rangle}{\Delta_p}&=&\frac{\Omega_a\Omega_b}{\Delta_p}.
\end{eqnarray}
Similarly, the large detuning condition $\Delta_{r}\gg\Omega_{c}$ permits us to eliminate the mediate state $|\chi\rangle$, then $H_I^1$ reduces to an equivalent form with two-atom Stark shifts of levels $|ee\rangle$ and $|rr\rangle$ and effective coupling between them
\begin{eqnarray}\label{A42}
&&\frac{\langle ee|H_I^1|\chi\rangle\langle \chi|H_I^1|ee\rangle}{\Delta_r}=\frac{2\Omega_c^2}{\Delta_r},\\
&&\frac{\langle rr|H_I^1|\chi\rangle\langle \chi|H_I^1|rr\rangle}{\Delta_r+\delta}\approx\frac{2\Omega_c^2}{\Delta_r},\\
&&\frac{\langle rr|H_I^1|\chi\rangle\langle \chi|H_I^1|ee\rangle}{\overline{\Delta}_r}\approx\frac{2\Omega_c^2}{\Delta_r}e^{i\delta t},\\
&&\frac{\langle ee|H_I^1|\chi\rangle\langle \chi|H_I^1|rr\rangle}{\overline{\Delta}_r}\approx\frac{2\Omega_c^2}{\Delta_r}e^{-i\delta t},
\end{eqnarray}
where $\delta\ll\Delta_r$ has been assumed and $1/\overline{\Delta}_r=[1/\Delta_r+1/(\Delta_r+\delta)]/2\approx1/\Delta_r$ \cite{james2007effective}. i.e.,
\begin{eqnarray}\label{A5}
H_{\rm eff}&=&\sum_{i=1}^2\frac{\Omega_a^2}{\Delta_p}|g\rangle_{i}\langle g|+\bigg(\frac{\Omega_b^2}{\Delta_p}
+\frac{\Omega_c^2}{\Delta_r}\bigg)|e\rangle_{i}\langle e| \cr
&&+\bigg[\frac{2\Omega_c^2}{\Delta_r}|ee\rangle\langle rr|e^{-i\delta t}+\sum_{i=1}^2\frac{\Omega_a\Omega_b}
{\Delta_p}|g\rangle_{i}\langle e|+{\rm H.c.}\bigg] \cr
&&+\frac{2\Omega_c^2}{\Delta_r}|rr\rangle\langle rr|.
\end{eqnarray}
The Stark shifts of ground states is  unwanted in our proposal, which can be canceled by other ancillary levels yielding opposite shifts of energy levels. After performing a rotating with respect to $U=\exp(i\delta t|rr\rangle\langle rr|)$,  Eq.~(\ref{A5}) is rewritten in the following time-independent form
\begin{equation}\label{A6}
H_{\rm eff}=\sum_{i=1}^2\Omega_{\rm eff}|g\rangle_{i}\langle e|+\lambda|ee\rangle\langle rr|+{\rm H.c.}+\Delta|rr\rangle\langle rr|,
\end{equation}
where $\Omega_{\rm eff}=\Omega_{a}\Omega_{b}/\Delta_{p}$, $\lambda={2\Omega_{c}^2/\Delta_{r}}$ and
$\Delta=\delta+{2\Omega_{c}^2/\Delta_{r}}$.
In order to further characterize the effective dynamics of system, we introduce the eigenstates of the two-atom transition Hamiltonian $(\lambda|ee\rangle\langle rr|+{\rm H.c.}+\Delta|rr\rangle\langle rr|)$ as follows
\begin{equation}\label{A7}
|\Psi_+\rangle=\cos\alpha|rr\rangle+\sin\alpha|ee\rangle,
\end{equation}
and
\begin{equation}\label{A8}
|\Psi_-\rangle=\sin\alpha|rr\rangle-\cos\alpha|ee\rangle,
\end{equation}
which correspond to eigenvalues $E_+=(\Delta+\sqrt{\Delta^2+4\lambda^2})/2$ and $E_-=(\Delta-\sqrt{\Delta^2+4\lambda^2})/2$, respectively, with $\alpha$=$\arctan[2\lambda/(\Delta+\sqrt{\Delta^2+4\lambda^2})]$. Through above steps, we recover the result of Eq.~(\ref{two04}). The derivation from Eq.~(\ref{two04}) to Eq.~(\ref{two06}) is straightforward as long as
the limiting conditions $R_1=|E_+/(\sqrt{2}\Omega_{\rm eff}\sin\alpha)|\gg1$ and $R_2=|E_-/(\sqrt{2}\Omega_{\rm eff}\cos\alpha)|\gg1$ are established. To better illustrate this process, we perform another rotating with respect to $\exp[-it(E_+|\Psi_+\rangle\langle\Psi_+|+E_-|\Psi_-\rangle\langle\Psi_-|)]$ on the basis of Eq.~(\ref{two04}) and obtain
\begin{eqnarray}\label{A9}
H_{\rm eff}&=&\sqrt{2}\Omega_{\rm eff}|gg\rangle\langle T^{'}|+\bigg(\sqrt{2}\Omega_{\rm eff}\sin\alpha e^{iE_+t}|\Psi_+\rangle \cr
&&-\sqrt{2}\Omega_{\rm eff}\cos\alpha e^{iE_-t}|\Psi_-\rangle\bigg) \langle T^{'}|+{\rm H.c.}.
\end{eqnarray}
where $|T\rangle=(|ge\rangle+|eg\rangle)/{\sqrt{2}}$. It can be seen clearly that the Hamiltonian of Eq.~(\ref{A9}) incorporates  the high-frequency oscillating terms proportional to $\exp{(iE_{\pm}t)}$, and these terms can be neglected while the resonant transition between states $|gg\rangle$ and $|T^{'}\rangle$ is preserved, hence a perfect ground-state blockade Hamiltonian of Eq.~(\ref{two06}) is achieved.
\bibliographystyle{apsrev4-1}
\bibliography{GroundblockadeV1}
\end{document}